\documentclass[epj]{svjour}
\usepackage{latexsym}
\usepackage{graphics}
\begin{document}
\title{Noise sensitivity of an atomic velocity sensor}
\subtitle{Theoretical and experimental treatment}
\author{Pierre~Clad\'e\inst{1}, Sa\"\i
da~Guellati-Kh\'elifa\inst{2}, Catherine~Schwob\inst{1}, Fran\c
cois~Nez\inst{1}, Lucile~Julien\inst{1} and Fran\c
cois~Biraben\inst{1}}
%
%
\institute{Laboratoire Kastler Brossel, \'Ecole Normale
Sup\'erieure, CNRS, UPMC, 4 place Jussieu, 75252 Paris Cedex 05,
France  \and CNAM-INM, Conservatoire National des Arts et
M\'etiers, 292 rue Saint Martin, 75141 Paris Cedex 03, France}
\date{Received: date / Revised version: date}
%
\abstract{We use Bloch oscillations to accelerate coherently
Rubidium atoms. The variation of the velocity induced by this
acceleration is an integer number times the recoil velocity due to
the absorption of one photon. The measurement of the velocity
variation is achieved using two velocity selective Raman $\pi$-
pulses: the first pulse transfers atoms from the hyperfine state
$5S_{1/2}$, $\left|F=2, m_F = 0\right>$ to $5S_{1/2}$,
$\left|F=1,m_F = 0\right>$ into a narrow velocity class. After the
acceleration of this selected atomic slice, we apply the second
Raman pulse to bring the resonant atoms back to the initial state
$5S_{1/2}$, $\left|F=2, m_F = 0\right>$. The populations in ($F=1$
and $F=2$) are measured separately by using a one-dimensional
time-of-flight technique. To plot the final velocity distribution
we repeat this procedure by scanning the Raman beam frequency of
the second pulse. This two $\pi$-pulses system constitutes then a
velocity sensor. Any noise in the relative phase shift of the
Raman beams induces an error in the
 measured velocity. In this paper
we present a theoretical and an experimental analysis of this
velocity sensor, which take into account the phase fluctuations
during the Raman pulses.
\PACS{
      {PACS-32.80.Pj}{Optical cooling of atoms; trapping}   \and
      {PACS-06.30.Gv}{Velocity, acceleration and rotation}
} 
} 
\authorrunning{P. Clad\'e {\it et al.}}
\titlerunning{Noise sensitivity of an atomic velocity sensor}
\maketitle
\section{Introduction}
\label{intro} The measurement of the recoil of an atom when it
absorbs a photon provides a way to determine the fine structure
constant $\alpha$ using atomic physics
\cite{Taylor,Wicht,Gupta,Battesti}. Since the first observation of
the recoil-induced spectral doubling in the $CH_4$ saturated
absorption peaks \cite{Hall}, only the development of atoms
cooling techniques renewed interest in measurement of the recoil
velocity $v_r$ (~$v_r = \hbar k /m$, where $k$ is the wave vector
of the photon absorbed by an atom of mass
$m$)\cite{Wicht,Gupta,Battesti}. The basic scheme of the photon
recoil, was previously proposed in reference \cite{Weiss} and a
simple version is illustrated in Fig.1: an atom in state
$\left|a\right>$, at rest in the laboratory frame, absorbs a
photon from rightward propagating laser beam with frequency
$\omega$.  The atom recoils by $\hbar k/m$ and the process has the
resonance condition deduced from energy conservation

\begin{figure}
\resizebox{0.45\textwidth}{!}{%
\includegraphics{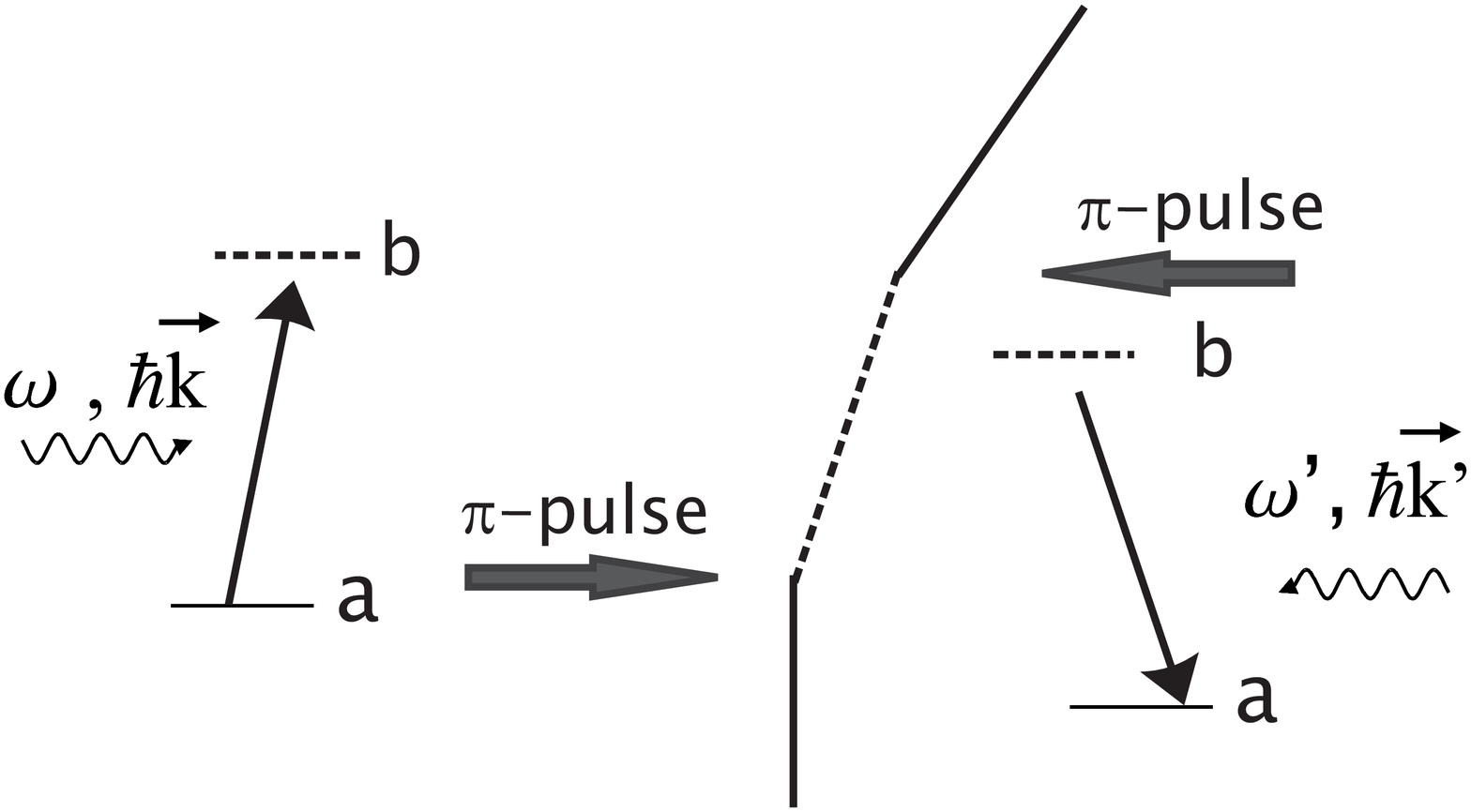} }
\caption{Basic way to measure the photon recoil: the atom jumps
from $|a>$ to $|b>$ by absorbing a rightward photon and acquires
one recoil, and then it goes back into $|a>$ by re-emitting a
leftward photon.
 }
\label{fig:1}       
\end{figure}

\begin{equation}
\omega_{ab}- \omega = \frac{\hbar k^2}{2 m} \label{eq1}
\end{equation}

The atom can be also de-excited from state $\left|b\right>$ by a
photon from a leftward propagating beam of frequency
$\omega^{\prime}$, the new resonance condition being

\begin{equation}
 \omega_{ab}-\omega^\prime  = - \frac{\hbar \overrightarrow{k}.
\overrightarrow{k}^{\prime}}{m}-\frac{\hbar k^{\prime 2}}{2 m}
\label{eq2}
\end{equation}

Thus, the two resonances are shifted relative to each other by

\begin{equation}
\omega - \omega^\prime = - \frac{\hbar (\overrightarrow{k} +
\overrightarrow{k}^\prime )^2}{2 m} \label{eq3}
\end{equation}

 If we fix $\omega$ and scan $\omega^\prime$ to find the maximum
number of atoms that come back to state $\left|a\right>$, we can
measure this frequency difference and hence deduce the recoil
shift. The ideal recoil measurement described above will be more
realistic using velocity-selective Raman transitions
\cite{Kasevich}. Transitions of this kind have two relevant
advantages: first the effective frequency is the hyperfine
splitting which is a microwave frequency and the effective
momentum kick is equal to that obtained with optical photons
(large Doppler shift). Second, as these transitions involve ground
state atomic levels, the linewidth of the stimulated transition,
and thus the width of the velocity distribution, is limited only
by the interaction time which is quite long when cold atoms are
used.

\begin{figure}
\resizebox{0.45\textwidth}{!}{%
\includegraphics{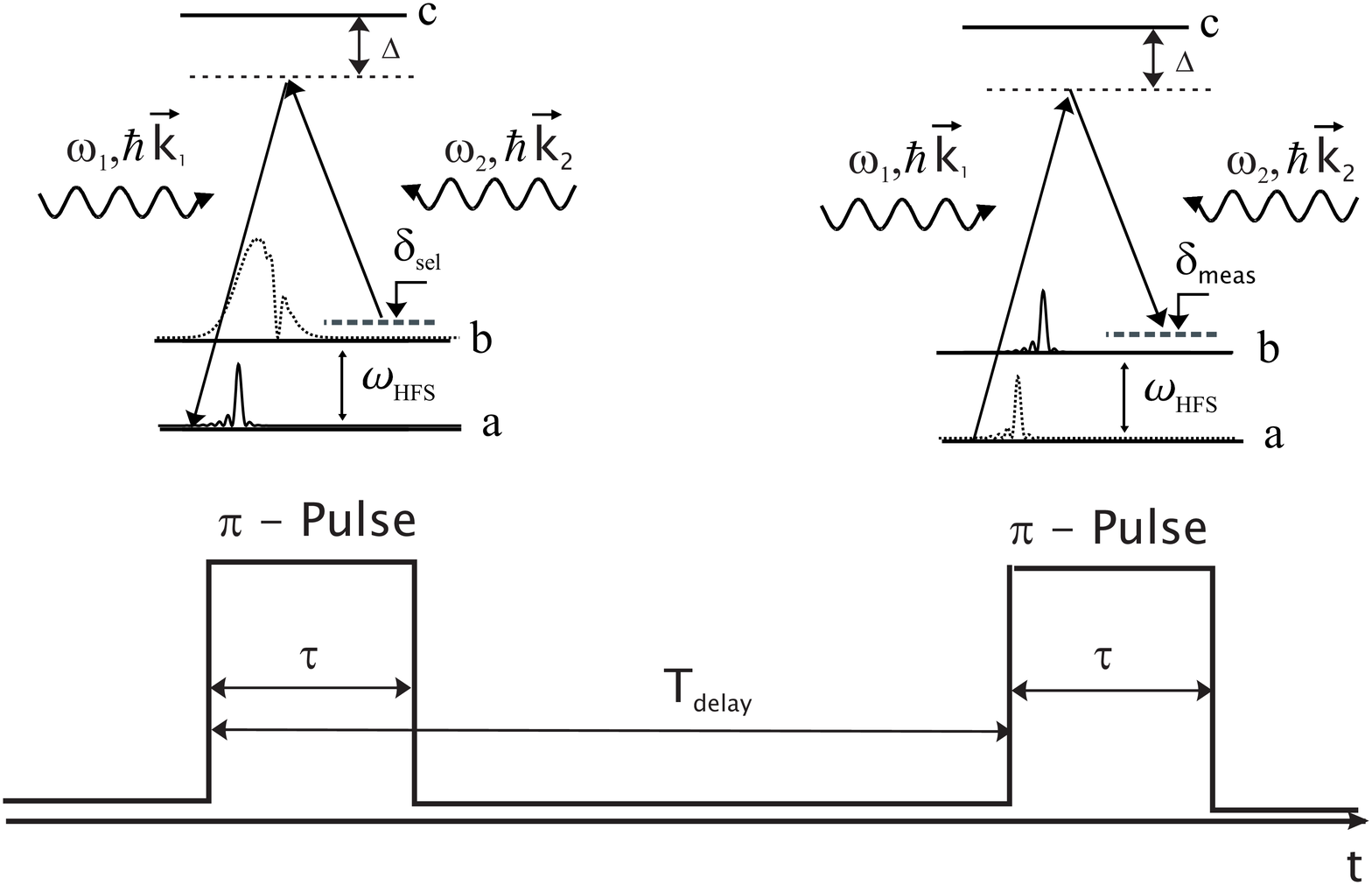} }
\caption{Principle of the velocity sensor, the first $\pi$-pulse
transfers a narrow velocity class from the level $\left|b\right>$
to the level $\left|a\right>$ (selection) and the second
$\pi$-pulse transfers the accelerated atoms back to the level
$\left|b\right>$ (measurement).}
\label{fig:2}       
\end{figure}

Let us consider an atomic cold sample where, after a laser cooling
process, the atoms, all in a well defined internal state
$\left|b\right>$, are illuminated successively by two
velocity-selective Raman $\pi$-pulses. The Raman excitation is
realized by two counter-propagating laser beams at frequencies
$\omega_1$ and $\omega_2$, and wave vectors \overrightarrow{k_1}
and \overrightarrow{k_2}. When the resonance condition:

\begin{equation}
\delta_{sel}= \omega_2 -
\omega_1-\omega_{HFS}=\overrightarrow{v_i}
(\overrightarrow{k_1}-\overrightarrow{k_2})+\frac{\hbar
(\overrightarrow{k_1} - \overrightarrow{k_2})^2}{2 m} \label{eq4}
\end{equation}

is fulfilled, the first $\pi-pulse$ transfers the atoms, in narrow
velocity class around the mean velocity $v_i$, from state
$\left|b\right>$ to $\left|a\right>$ (see Fig.\ref{fig:2}). Here
$\delta_{sel}$ is the detuning of the co-propagating Raman
transition.

After an acceleration which changes the mean velocity of the
atomic velocity class from $v_i$ to $v_f$, we apply a second
$\pi$-pulse and we shift the detuning to $\delta_{meas}$ so as we
satisfy the resonance condition (equation (\ref{eq4})) for the
mean velocity $v_f$. By scanning the detuning $\delta_{meas}$ of
the final Raman pulse to get maximum of atoms back into initial
state $\left|b\right>$, we determine the variation of velocity
$\Delta\overrightarrow{v}$ by

\begin{equation}
\Delta\overrightarrow{v}\cdot
(\overrightarrow{k_1}-\overrightarrow{k_2})=
(\delta^{max}_{meas}-\delta_{sel}) \label{eq5}
\end{equation}

This system constitutes a velocity sensor. In our experiment the
atoms are coherently accelerated using Bloch oscillations in a
periodic optical potential \cite{Battesti,Ben Dahan,Peik}. In this
case, the  velocity variation of the center of mass is an integer
times the recoil velocity $ v_r$. In this paper we shall ignore
this intermediate step and only focus on the study of the velocity
sensor described above. In the following we investigate
theoretically the number of atoms in the state $\left|b\right>$
after the second $\pi$-pulse, starting by the determination of the
Raman transition probability and  taking into account the relative
phase noise between the two counter-propagating beams. We then
calculate the noise sensitivity of the velocity sensor and the
ordinary variance of the measured atoms. Finally, we present the
experimental set-up and discuss how the experimental compares with
our theoretical model. We underline that previously other groups
have studied the phase fluctuations of the Raman beams in atom
interferometers \cite{Peters,leduc}. The originality of this work
is to take into account the effects of the phase fluctuations
during the Raman pulses and not only between the pulses.

\section{Theory}

The theory of velocity-selective stimulated Raman transitions was
been widely studied by \cite{Kasevich,Moler}. In the subsection
2.1, we investigate the stimulated Raman transition probability
considering the relative phase noise $\varphi(t)$ (time
dependence) between the two beams. In subsection 2.2, we consider
the double $\pi$-pulse and we determine the fraction of atoms at a
given detuning $\delta$ of the Raman beam frequency. We then
deduce the sensitivity of the velocity sensor by expressing the
ordinary variance as function of a power spectral density of the
phase noise.

\subsection{One pulse Raman transition}
\label{sec:2}

We consider an atom that has a level scheme shown in
Fig.\ref{fig:2}. with a ground state hyperfine interval
$\omega_{HFS}$. This atom is irradiated, along the $z$ axis, by
two counter-propagating laser beams ( $\omega_1$,
\overrightarrow{k_1}) and ($\omega_2$, \overrightarrow{k_2}).

The states $|a,p-\hbar k_1\rangle$ and $|b,p+\hbar k_2\rangle$ are
coupled to $|c, p\rangle$ respectively by the wave ($\omega_1$,
\overrightarrow{k_1}) and ($\omega_2$, \overrightarrow{k_2}). The
atomic system is then equivalent to a two-level system coupled by
a two-photon transition with an effective Rabi frequency :

\begin{equation}
\Omega= \frac{\Omega_1^* \Omega_2}{2\Delta}
\label{eq6}
\end{equation}

where $\Delta=\omega_1-\omega_{ac}\approx\omega_2-\omega_{bc}$ is
the one photon detuning (see Fig.\ref{fig:2}) and the Rabi
frequencies $\Omega_1$ and $\Omega_2$ are defined by

\begin{equation}
\Omega_1 = - \frac{\langle
a|\overrightarrow{d}.\overrightarrow{E_1}|c\rangle}{2\hbar},
\Omega_2 = - \frac{\langle
b|\overrightarrow{d}.\overrightarrow{E_2}|c\rangle}{2\hbar}
\label{eq7}
\end{equation}

$\overrightarrow{E}_n$, ($n=1,2$) is the electric field of the
travelling wave $n$,
 $\overrightarrow{d}$ is the electric dipole operator.

To include the relative phase noise $\varphi(t)$ between the two
Raman beams, we express the effective Rabi frequency as

\begin{equation}
\Omega(t) = \Omega_0  e^{i\varphi (t)}
\label{eq8}
\end{equation}

Assuming that $\varphi(t) \ll 1$, the Hamiltonian of this
two-level system can be linearized as the sum of $H_0$ and
$H_{pert}$, where in convenient Pauli matrix representation

\begin{equation}
H_0 = \hbar \left(\frac{\delta}{2} \sigma_z + \frac{\Omega_0}{2}
\sigma_x \right) \label{eq9}
\end{equation}

$\delta$ is the detunning of $\omega_1-\omega_2$ from the
transition $|a,p-\hbar k_1\rangle \longrightarrow |b,p+\hbar
k_2\rangle$.

The time dependent perturbative hamiltonian in first order
approximation is given by

\begin{equation}
 H_{pert}(t)= i\hbar \frac{\Omega_0}{2} \varphi(t) \sigma_x
\label{eq10}
\end{equation}

The state of a quantum system at a final time $t_f$ is related to
its state at an earlier time $t_i$ via the evolution operator U

\begin{equation}
|\psi (t_f)\rangle = U(t_f, t_i)|\psi (t_i)\rangle \label{eq11}
\end{equation}

using the time dependent perturbation theory, in first order, the
evolution operator U is given by

\begin{equation}
U(t_f - t_i)= U_0 (t_f - t_i) + \frac{1}{i\hbar}\int_{t_i}^{t_f}
{U_0 (t_f - t) H_{pert}(t) U_0 (t - t_i) dt} \label{eq12}
\end{equation}

where
\begin{equation}
U_0(t) = e^{-i \frac{H_0 t} {\hbar}} \label{eq13}
\end{equation}

The time dependent transition probability $P$ from level
$|a\rangle$ to level $|b\rangle$ is

\begin{equation}
P(\delta)= |\langle a |U | b \rangle|^2
\label{eq14}
\end{equation}

Substituting the relations (\ref{eq12}) and (\ref{eq13}) into
equation (\ref{eq14}) we show that the transition probability can
be written as

\begin{equation}
 P(\delta)=P^0(\delta)+P^1(\delta)
\label{eq15}
\end{equation}

$P^0$ is given by the Rabi formula:

\begin{equation}
P^0(\delta) = \frac{\Omega_0^2}{\Omega^{\prime
2}}\sin^2\frac{\Omega^{\prime} (t_f - t_i) }{2} \label{eq16}
\end{equation}

and $P^1$, the time dependent transition probability to first
order in the relative phase noise is given by

\begin{equation}
P^1(\delta) = -\delta\frac{\Omega_0^2}{\Omega^{\prime 2}} sin
\frac{\Omega^\prime(t_f - t_i)}{2}
\int_{t_i}^{t_f}{\varphi(t)sin\frac{\Omega^\prime (2t-t_f -
t_i)}{2} dt} \label{eq17}
\end{equation}
 where
\begin{equation}
\Omega^\prime = \sqrt{\Omega_0^2+\delta^2} \label{eq18}
\end{equation}
is the generalized Rabi frequency.

\subsection{Selection and Measurement}

We consider now an atom in internal state $|b\rangle$ with an
initial velocity $v_i$ along the beams axis. This atom is
illuminated consecutively by two Raman $\pi$-pulses with the same
duration $\tau$ and separated by the time interval $T_{delay}$
(see Fig.\ref{fig:2}). During the time interval between the two
$\pi$-pulses, the atom is accelerated to change its velocity by
$\Delta v$ (the final velocity  is then $v_f=v_i+ \Delta v$).
$P_{sel}(\delta_{sel} - 2k v_i)$ and $P_{meas}(\delta_{meas} - 2k
v_f)$ are respectively the probability to make the first and the
second Raman transition.

The experimental proceeding of the velocity sensor was described
in the first section and illustrated in the Fig.\ref{fig:2}. The
atoms remaining in level $|b\rangle$ after the first $\pi$-pulse,
are pushed away using a resonant laser beam. The distribution
velocity of the selected velocity class is supposed constant along
the width of the selection ($n(v)=n_0$) (in fact, the typical
width of the initial distribution obtained with an optical
molasses in a few recoils, whereas the first $\pi$-pulse selects
atoms in a velocity class of about $v_r/30$). After the second
pulse, we measure separately the number of atoms in state
$|a\rangle$ and $|b\rangle$ using two parallel, horizontally
propagating probe beams, placed 15 cm below the center of the trap
and separated vertically by 1 cm. The number $N_b$ of atoms
transferred by the second pulse is equal to the contribution of
all selected atoms weighted by the probability to make the second
$\pi$-pulse Raman transition:

\begin{eqnarray}
N_b (\delta_{meas}-\delta_{sel}) = \nonumber\\\frac{n_0}{2k}
\int_{-\infty}^{+\infty}{P_{sel} (\delta_{sel} + \eta) P_{meas}
(\delta_{meas}-2k \Delta v +\eta) d\eta} \label{eq19}
\end{eqnarray}

where $\eta = - 2k v_i$.

The total number $N_a + N_b$ of atoms detected after the second
pulse is nothing more than the number $N_{sel}$ of atoms selected
by the first $\pi$-pulse:

\begin{equation}
N_{sel}(\delta_{sel})=n_0
\int_{-\infty}^{\infty}P_{sel}(\delta_{sel}-2kv_i)dv_i
\end{equation}

To eliminate the fluctuations of the initial number of atoms, we
consider in the following the probability $\mathcal{P}=
N_b/(N_a+N_b)$ which represents the velocity distribution of the
measured atomic fraction. By inserting (\ref{eq15}) in
(\ref{eq19}) and using the fact that $P^1$ is an even function, we
finally obtain the correction of $\mathcal{P}$ to first order in
$\varphi(t)$:

\begin{equation}
\mathcal P^1(\delta + 2k\Delta v)= \frac{
\int_{-\infty}^{+\infty}{P^0 ( \eta-\delta) (P_{sel}^1 (\eta)-
P_{meas}^1 (\eta)) d\eta}}{\int_{-\infty}^{\infty}P^0(\eta)d\eta}
\label{eq20}
\end{equation}

where in thus case $\delta$ is equal to $\delta_{meas}-2k\Delta v-
\delta_{sel}$.

\subsection{Determination of the transfer function $H(f,\delta)$}

The best way to test the propagation of the phase fluctuation
$\varphi(t)$ on the velocity sensor is to calculate the ordinary
variance $\sigma_{\mathcal P}$ of the probability $\mathcal{P}$ to
make the two Raman transitions.

\begin{equation}
\sigma_{\mathcal{P}}^2(\delta) = <(\mathcal{P}- <\mathcal{P}>)^2>
\label{eq20a}
\end{equation}

The probability $\mathcal P$ is a linear function of $\varphi (t)$
(inserting (\ref{eq17}) in (\ref{eq20})). Assuming that $\varphi$
is a stationary random variable, we can express $\sigma_{\mathcal
P}$ as a function of the density of the noise $\Phi_f$

\begin{equation}
\sigma_{\mathcal{P}}^2 (\delta) =
\int_{-\infty}^{+\infty}{\Phi_f^2~~H^2 (f, \delta) df }
\label{eq24}
\end{equation}

where

\begin{equation}
\Phi_f^2 = 4~\int_{-\infty}^{+\infty} d\tau e^{2\pi i f\tau}
\langle\varphi(t+\tau) \varphi(t)\rangle \label{eq24b}
\end{equation}

and $H(f,\delta)$ represents the transfer function or the noise
sensitivity of the velocity sensor. To easily calculate this last
function using (\ref{eq20}), we will assume that the phase
fluctuation between the Raman beams can be expressed as

\begin{equation}
\varphi(t)=\sum_f \Phi_f\sqrt{\Delta f}\cos(2\pi f t +\varphi_f)
\label{eq22}
\end{equation}

where $\varphi_f$ are arbitrary phases at each frequency $f$ (we
assume that the phases $\varphi_f$ between two different
frequencies are independent). At the limit where the frequency
band $\Delta f \rightarrow 0$, the two points of view in equations
(\ref{eq24b}) and (\ref{eq22}) give the same result for
$H(f,\delta)$. In equation (\ref{eq22}) the noise density $\Phi_f$
is expressed in $(rad/\sqrt{Hz})$.

First we calculate the one $\pi$-pulse transition using the
expression of $\varphi(t)$ (\ref{eq22}) in (\ref{eq17}):

\begin{eqnarray}
P^1(\delta)= -\delta \frac{\Omega_0^2}{\Omega^{\prime
2}}sin\frac{\Omega^\prime}{2}(t_f - t_i) \sum_f\Phi_f  \sin(\pi f
(t_f+t_i)+\varphi_f) \nonumber\\
\left(\frac{sin(( 2\pi f +\Omega^\prime)\frac{(t_f-t_i)}{2})}{2\pi
f +\Omega^\prime}- \frac{sin(( 2\pi
f -\Omega^\prime)\frac{(t_f-t_i)}{2})}{2\pi f -\Omega^\prime}\right)\sqrt{\Delta f}\nonumber\\
\label{eq23}
\end{eqnarray}

Second we calculate the two Raman transitions probability (for two
$\pi$-pulses) substituting $P^1$  by (\ref{eq23}) in (\ref{eq20})

\begin{equation}
\mathcal P^1(\delta + 2k\Delta
v)=\sum_f\Phi_f~h(f,\delta)~\cos(\pi f (T_{delay}+\tau)+\varphi_f)
\sqrt{\Delta f} \label{eq25a}
\end{equation}

where

\begin{eqnarray}
h(f,\delta) =\int_{-\infty}^{+\infty}2 P_0( \eta-\delta) \delta
\frac{\Omega_0^2}{\Omega^{\prime 2}} \sin\frac{\Omega^\prime
\tau}{2} \sin(\pi f
T_{delay})\nonumber\\
\left(\frac{sin( 2\pi f -\Omega^\prime)\frac{\tau}{2}}{2\pi f
-\Omega^\prime}- \frac{sin(2\pi f
+\Omega^\prime)\frac{\tau}{2}}{2\pi f +\Omega^\prime}\right) d\eta
\label{eq25}
\end{eqnarray}

with $\tau = t_f-t_i$ and $T_{delay}$ is the time interval between
the two $\pi$-pulses. To simplify the presentation of the formula
(\ref{eq25}) the normalization factor in (\ref{eq20}) is omitted.

Since for each frequency, $\varphi_f$ is a random variable with an
uniform distribution on $[0, 2\pi]$, then
\begin{equation}
<\mathcal{P}^1> = 0~~and~~<(\mathcal P^1)^2>=\sum
\frac{1}{2}~\Phi_f^2~h^2(f,\delta) \Delta f \label{eq25c}
\end{equation}

Substituting (\ref{eq25c}) in the definition of the ordinary
variance $\sigma_{\mathcal P}$, we deduce the expression of the
transfer function $H(f,\delta)$
\begin{equation}
H(f,\delta)=\frac{1}{\sqrt{2}}~|h(f,\delta)| \label{eq25d}
\end{equation}

This function depends on the pulse interval on $sin (\pi f
T_{delay})$ (see (\ref{eq25})): for each $T_{delay}$ there are
certain frequencies at which the phase noise does not have any
effect.

\section{Experiment}

\begin{figure}
\resizebox{0.45\textwidth}{!}{%
\includegraphics{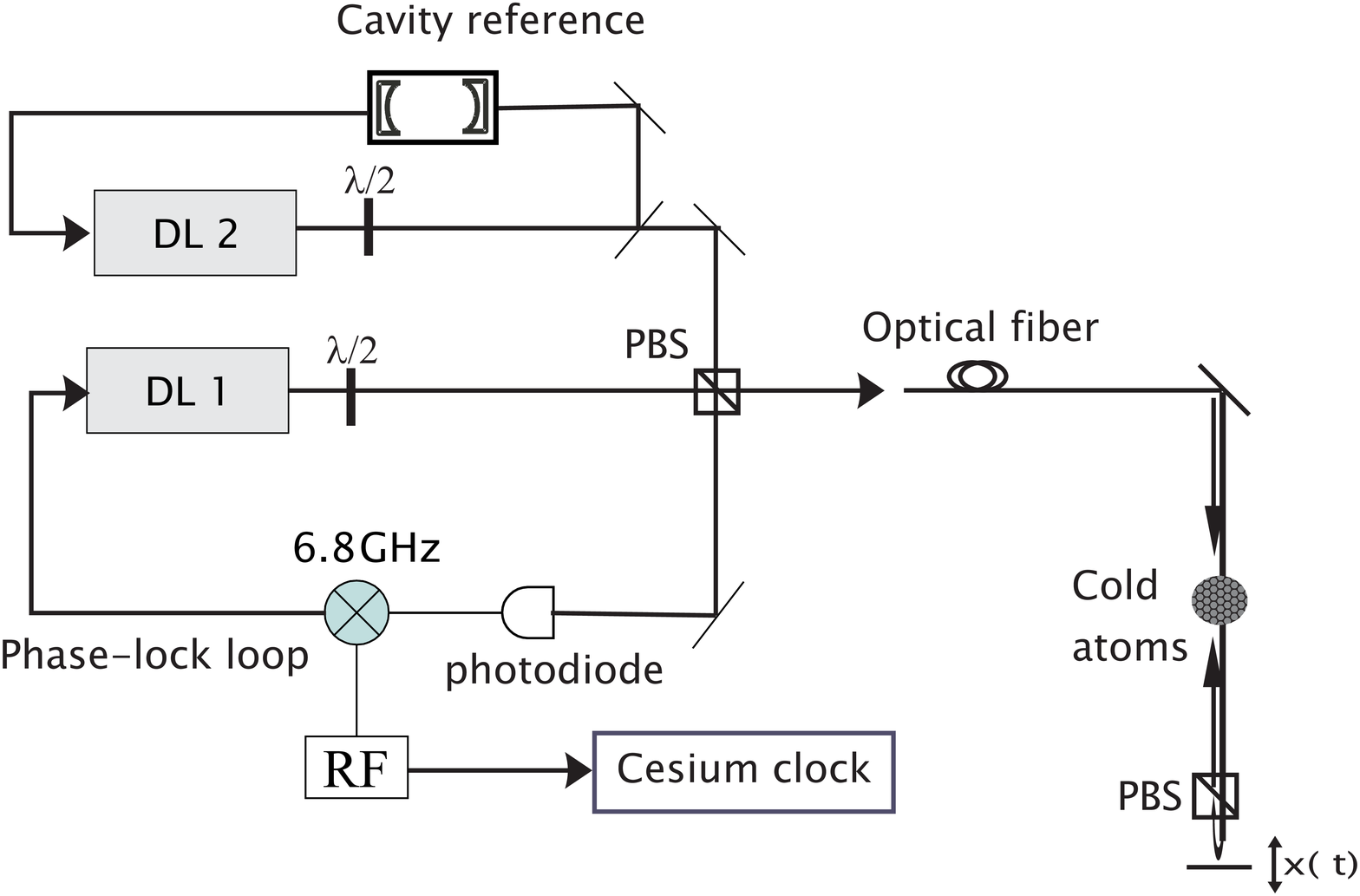} }
\caption{Experimental setup of the velocity sensor: The two laser
diodes are phase locked using a tunable microwave chain. The two
Raman beams with lin$\perp$lin polarizations are injected into the
same optical fiber. After passing through the vacuum chamber only
the beam $k_2$ is retroreflected by the horizontal mirror allowing
a counter-propagating excitation.}
\label{fig:3}       
\end{figure}

\begin{figure}
\resizebox{0.45\textwidth}{!}{%
\includegraphics{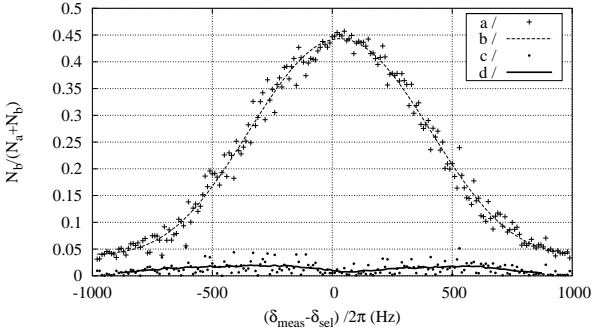} }
\caption{a) Final velocity distribution of atoms in hyperfine
state $|F=2, m_F=0\rangle$ measured using two counter-propagating
Raman beams and normalized to $N_{sel}= N_a+N_b$ the number of
atoms selected by the first pulse ($N_a$ and $N_b$ are
successively the number of atoms measured in the hyperfine states
$|F=1, m_F=0\rangle$ and $|F=2, m_F=0\rangle$. b) Theoretical fit.
c) Difference between the experimental data and the theoretical
fit. d) Smoothing curve of the last data using a fixed window. We
note finally that in x-axis the velocity is expressed in terms of
frequency.
\label{fig:4}}       
\end{figure}

An optical molasses loaded by a 3-D magneto-optical trap provides
a cold $^{87}Rb$ atomic sample \cite{Battesti}. For the initial
selection and the final measurement, the two Raman beams are
generated by two laser diodes injected by two grating-stabilized
extended-cavity laser diodes (ECLs). A fast photodiode and a
tunable RF frequency chain are used to phase lock one ECL on the
other one. The two beams have linear orthogonal polarizations.
After passing through the vacuum cell, one beam is retroreflected
by a horizontal mirror (see Fig.\ref{fig:3}). A typical scan of
final velocity distribution of Rubidium atoms transferred by the
second pulse from $5S_{1/2}$ $\left|F=1, m_F = 0\right\rangle$ to
$5S_{1/2}$ $\left|F=2,  m_F = 0\right\rangle$, is shown in Fig.
(\ref{fig:4}.b). The noise level affecting a measured spectrum is
not uniform: it is lightly greater on the slopes than on the top.
A better illustration of the noise distribution affecting the
spectrum can be achieved by plotting the difference between the
theoretical fit and the experimental data (Fig.\ref{fig:4}). Thus
is a proof that the spectral noise is not yet dominated by the
atomic shot noise (number of detected atoms) but by the Raman
phase noise. This phase noise can arise from optical noise (laser
noise, fiber noise, phase lock noise, ...) or the vibration noise
of the retroreflection mirror (indeed, the velocity of the atoms
is measured in the frame of this mirror).

The resolution of our velocity sensor is then mainly limited by
the Raman phase noise. In the next section we analyze the
experimental results using the theoretical model developed above
considering only the vibrational noise affecting the
retroreflection mirror.

\section{Analysis of the experimental results}

The Raman beam phase noise includes different noise sources, it
can be written essentially as a sum of two contributions:
\begin{equation}
\varphi(t) = [\varphi_1 (t) - \varphi_2 (t)] -2k_2 x(t)
\label{Bruit} \label{eq21}
\end{equation}
where $\varphi_i (t)$ is the optical phase of the beam $i$, $x(t)$
characterizes the motion  of the retroreflection mirror.

\begin{figure}
\resizebox{0.45\textwidth}{!}{%
\includegraphics{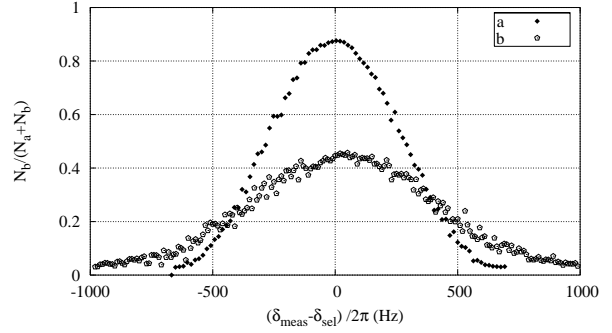} }
\caption{The fraction of atoms transferred by the second Raman
$\pi$-pulse: a) Co-propagating Raman beams configuration, b)
Counter-propagating configuration. In this last case, the
Doppler-sensitive Raman transition is performed only for a
resonant velocity class. This explain the amplitude and the FWHM
difference between the two spectra.}
\label{fig:5}       
\end{figure}

A straightforward way to distinguish between the vibration noise
and other phase noises, is to compare the Doppler-insensitive
Raman transition spectrum obtained using a co-propagating beams
(unaffected by vibrations of the retroreflection mirror)
(Fig.\ref{fig:5}.a) to the Doppler-sensitive Raman transition
spectrum driven by the counter-propagating laser beams
(Fig.\ref{fig:5}.b).

\begin{figure}
\vspace{5cm}
\resizebox{0.45\textwidth}{!}{%
\includegraphics{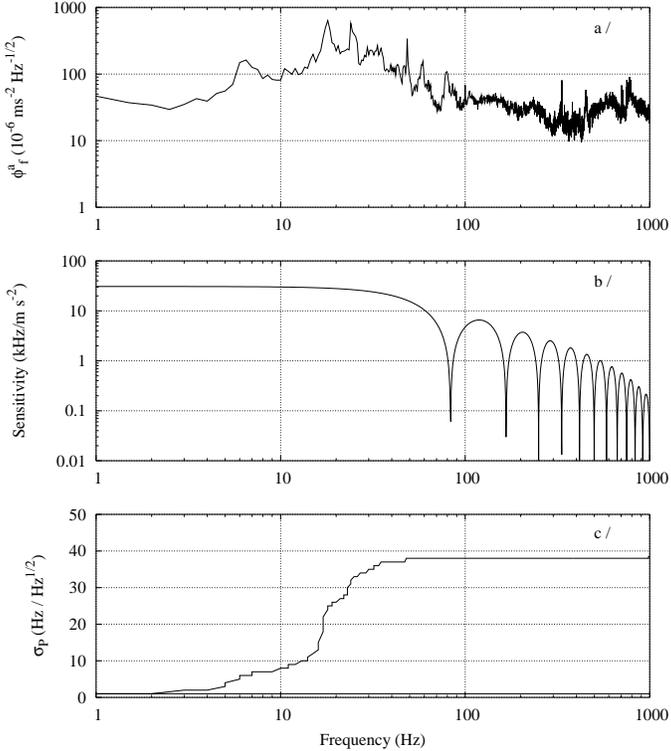} }
\caption{(a). The acceleration noise spectral density deduced from
the vibrational spectrum of the retroreflection mirror measured by
an accelerometer. (b) The theoretical velocity sensor noise
sensibility for pulse duration $\tau$=1~ms and pulse interval
$T_{delay}$=12~ms. (c) Predicted variance of the atoms fraction
transferred by the second pulse integrated up to a certain
frequency calculated using in the formula (\ref{eq24}) the
measured phase noise spectral density.}
\label{fig:6}       
\end{figure}

This illustration shows that the relative noise is more than one
order of magnitude lower in Fig.\ref{fig:5} than in the case of
the Doppler-sensitive Raman transition. Given thus, the optical
phase noise $\varphi_1(t)-\varphi_2(t)$ is not a relevant noise in
our experimental set-up. In order to test the theoretical model
presented above, we only take into account in expression
(\ref{eq21}) the vibration term. The phase noise spectral density
can be expressed \cite{note1} as

\begin{equation}
\Phi_f = \frac{2 k}{(2\pi f)^2}\Phi_f^{a} \label{eq27}
\end{equation}
where $\Phi_f^{a}$ is the acceleration noise spectral density,
deduced from the acceleration of the mirror which is measured by a
low-noise, low-frequency accelerometer (IMI Sensors-626A). The
Fig.\ref{fig:6}.a, shows the acceleration noise power spectrum
($\Phi_f^{a}$) of the retroreflection mirror. It is determined
using a numerical Fourier transform of the monitored accelerometer
signal. The rms value of the vibrational phase noise integrated on
the pulse duration is estimated to $0.1~rad$, and remains in the
validity range of the perturbative approach used in our
theoretical model.

The vibration sensitivity of the velocity sensor ($2kH(f)/(2\pi
f)^2$) is plotted for a pulse duration of $1$~ms and a time
spacing pulse $ T_{delay}$ of $12$ ms using the
(Fig.\ref{fig:6}.b). This curve shows that the velocity sensor
acts as a low-pass filter of vibrations, with a cut off frequency
of about 35 Hz. The effect of the mechanical vibration on the
uncertainty of the velocity measurement can be illustrated by
plotting a predicted variance $\sigma_{\mathcal P}$ using the
acceleration noise spectral density (Fig.\ref{fig:6}.a) and the
vibration sensitivity (Fig.\ref{fig:6}.b). It appears that the
main part of the vibration noise in our experimental set-up comes
from frequencies between 10 and 30~Hz (Fig.\ref{fig:6}.c).

\begin{figure}
\resizebox{0.45\textwidth}{!}{%
\includegraphics{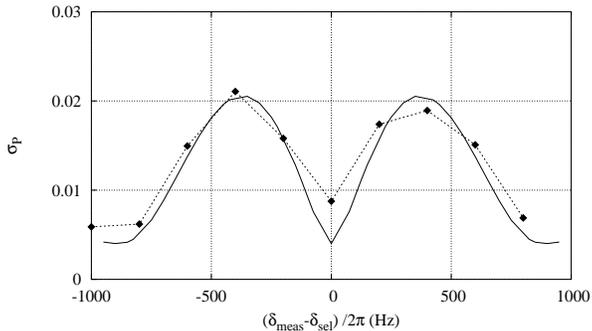} }
\caption{Variance on the fraction of atoms transferred by the
second $\pi$-pulse, dashed line experimental result, solid line
predicted value.}
\label{fig:7}       
\end{figure}

As predicted by the theoretical model and illustrated in the
typical velocity distribution spectrum (Fig.\ref{fig:4}), the
noise of the velocity sensor depends on the Raman detuning
$\delta$. By making several measurements at the same detuning
$\delta$, we measure the statistical variance $\sigma_{\mathcal
P}$ of the transition probability of the two Raman pulses
(Fig.\ref{fig:7}).

The good qualitative and quantitative agreements with the
predicted variance, allow us to confirm that the theoretical model
developed in this paper is a powerful tool for quantifying and
hence controlling  the different noises of the Raman beams.

\begin{figure}
\resizebox{0.45\textwidth}{!}{%
\includegraphics{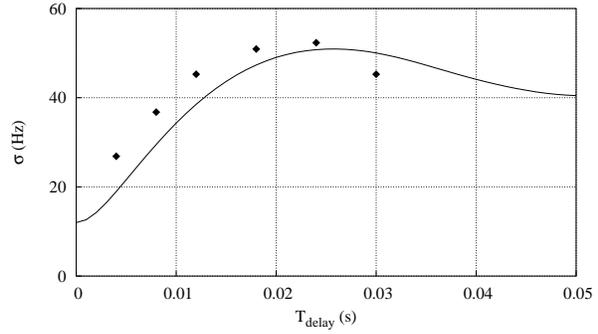} }
\caption{The uncertainty of the measured velocity expressed in
term of frequency, for different $T_{delay}$, predicted value
(line) and experimental value (dot). }
\label{fig:8}       
\end{figure}

The time interval $T_{delay}$ is a critical parameter of the
experiment, it determines the number of additional recoils
transferred by the Bloch oscillations process and hence the
resolution of the photon recoil measurement. It will be useful to
understand how this parameter operates on the uncertainty. In the
Fig.\ref{fig:8}, the dots present the uncertainty on the measured
velocity in term of frequency. This uncertainty is deduced from
the least-square fit of the experimental data points of the final
velocity distribution by the non-perturbative part of the two
pulses transition probability ${\mathcal P}^0$. We remind that
this probability is determinated substituting in (\ref{eq19})
$P_{sel}$ and $P_{meas}$ by the one pulse Raman non-perturbative
transition probability $P^0$ defined in (\ref{eq16}). To predict
this uncertainty, denoted $\sigma$, using the previous model we
use the following formula

\begin{equation}
\sigma^2 =\frac{1}{n} \frac{\sum_\delta \sigma_{\mathcal
P}^2}{\sum_\delta(\frac{\partial{\mathcal
P^0}}{\partial{\delta}})^2}
 \label{eq28}
\end{equation}

where $n$ is the number of the sample. This expression is obtained
by substituting in the expression of the uncertainty given by a
least square fit algorithm, the deviation of the numerical data
from the theoretical function, by the theoretical mean uncertainty
$\sigma_{\mathcal P}$. In this plot the noise increases with
$T_{delay}$ and reaches a maximum value, the noise decreases then,
because the band pass of the velocity sensor varies as
$1/T_{delay}$ and then it filters the high vibrational noise
frequencies.

\section{Conclusion}

In this paper we have developed a simple theoretical tool, to
characterize the noise of an atomic velocity sensor. We have
focused on the phase fluctuations of the Raman beams during the
pulse, such effects are very important in our non-interferometric
velocity sensor where the resolution is inversely proportional to
the Raman pulses duration. The experimental illustration was here
limited to the vibrational noise, but the model can be used for
any other phase noise at the limit of the validity of the
perturbative approach. This tool allows us to understand how to
implement the experimental improvements, essentially the vibration
isolation.

\begin{acknowledgement}
We thank A. Clairon and co-workers for valuable discussions. This
experiment is supported in part by the Bureau National de
M\'etrologie (Contrats 993009 and 033006) and by the R\'egion Ile
de France (Contrat SESAME E1220).
\end{acknowledgement}


\begin{thebibliography}{}
\bibitem{note1} The phase noise induced by the vibration of the retroreflecting mirror
is $\varphi(t)= 2 k_2 x(t)$ then the acceleration of this mirror
$a(t)$ can be written as $a(t)= \sum \frac{(2\pi f)^2 \Phi_f}{2k}
\sqrt{\Delta f} \cos(2\pi f t +\varphi_f)$.
\end{thebibliography}
\end{document}